# Crystal Structures and Phase Transitions of the van-der-Waals Ferromagnet VI₃

P. Doležal[1], M. Kratochvílová[1], V. Holý[1], P. Čermák[1], V. Sechovský[1], M. Dušek[2], M. Míšek[2], T. Chakraborty[3], Y. Noda[4], Suhan Son[3,5], Je-Geun Park[3,5]

[1]*Charles University, Faculty of Mathematics and Physics, Department of Condensed Matter Physics, Ke Karlovu 5, 121 16 Prague 2, Czech Republic*

[2]*Institute of Physics, Academy of Sciences of Czech Republic, v.v.i, Na Slovance 2, 182 21 Prague 8, Czech Republic*

[3]*Center for Correlated Electron Systems, Institute for Basic Science, Seoul 08826, Korea*

[4]*Institute of Multidisciplinary Research for Advanced Materials, Tohoku University, Sendai 980-8577, Japan*

[5]*Department of Physics and Astronomy, Seoul National University, Seoul 08826, Korea*

The results of a single-crystal X-ray-diffraction study of the evolution of crystal structures of VI₃ with temperature with emphasis on phase transitions are presented. Some related specific-heat and magnetization data are included. The existence of the room-temperature trigonal crystal structure R-3 (148) has been confirmed. Upon cooling, VI₃ undergoes a structural phase transition to a monoclinic phase at $T_s \sim 79$ K. $T_s$ is reduced in magnetic fields applied along the trigonal $c$-axis. When VI₃ becomes ferromagnetic at $T_{FM1} \sim 50$ K, magnetostriction-induced changes of the monoclinic-structure parameters are observed. Upon further cooling, the monoclinic structure transforms into a triclinic variant at 32 K which is most likely occurring in conjunction with the previously reported transformation of the ferromagnetic structure. The observed phenomena are preliminarily attributed to strong magnetoelastic interactions.

Van der Waals ferromagnets have recently been subject of intensive research due to their potential use in atomically thin devices with spintronic and optoelectronic functionalities [1-4]. Ferromagnetism persisting down to the monolayer limit is an ingredient essential for future spintronic applications. This was probably the main reason why $CrI_3$ with the highest Curie temperature ($T_C$ = 61 K) among the transition metal trihalides ($TX_3$) has attracted so much research interest lately [4-13]. On the other hand, $VI_3$, also known already since the '60s [14-16], has received significant attention only since this year [18-21]. $VI_3$ is also ferromagnetic (FM) with $T_C \sim$ 50 K, slightly lower than $CrI_3$.

The $TX_3$ trihalides are frequently dimorphic. Similar to $CrCl_3$ and $CrBr_3$, the low-temperature (LT) phase of $CrI_3$ is trigonal $BiI_3$-type with space group $R$-$3$ (148) while the high-temperature (HT) structure is monoclinic $AlCl_3$-type ($C2/m$ (12)). A first-order structure transition between the two crystallographic phases in $CrCl_3$, $CrBr_3$ and $CrI_3$ occurs at ~ 210, 420 and 240 K, respectively [5,21].

Much less is known about the crystal structures of the analogous $VX_3$ compounds. The only unambiguous information available is that, at room temperature (RT), $VCl_3$ and $VBr_3$ also possess the rhombohedral structure $R$-$3$ (148) [23-25]. Papers on $VI_3$, rapidly appearing within a short period, have provided contradictory statements. Son et al. [18] report the RT trigonal structure $P$-$31c$ (163), a structural phase transition at $T_s$ = 79 K and monoclinic crystal symmetry $C2/c$ (17). Kong et al. [19] have determined the trigonal structure $R$-$3$ (148) at 100 K and suggest a subtle structural phase transition at 78 K. In contrast, Tian et al. [20] claim that the structural phase transition of $VI_3$ is analogous to the structural transition of $CrI_3$, i.e. between the HT monoclinic structure $C2/m$ (15) (determined at 100 K) and the LT rhombohedral structure $R$-$3$ (148) determined at 60 and 40 K).

The present paper is focused on results of a study of the crystal structures and structural phase transitions in $VI_3$ by means of X-ray single-crystal diffraction (XRSCD) methods at temperatures between 300 K and 5 K, complemented by measurements of specific heat and magnetization as function of temperature ($T$) and magnetic field ($H$).

The refinement of the RT $VI_3$ crystal structure by the analysis of 792 independent diffraction peaks collected at 250 K provides the trigonal $R$-$3$ (148) space group as the best solution. Two possible structure models are suggested in literature: a) $R$-$3$ (148) [14,19] or b) P-31c (163) [18]. The present refinement of the structure with the space group $R$-$3$ (148) quickly converged with excellent $R$ values (see Table S1 in Supplemental Material [26]) with refined twin-volume fractions. Because some publications report crystal structures of this material based on the Laue symmetry -31m, refined successfully from powder data, we tried to ignore the R centering and attempted to solve the structure using space group $P$-$31m$ (163). Although the refinement was very unstable and the ADP parameters of iodine atoms were mostly wrong, the resulting R-value was about 0.06. This demonstrates that even an incorrect structure may provide quite a good fit in Rietveld refinement. In this case the accuracy is considerably lower compared with structure determination from single-crystal diffraction data. The detailed description of the RT experiment is presented in Supplemental Material (SM) [26].

The LT XRSCD measurements evidence that the crystal symmetry is lowered from the HT trigonal symmetry of $VI_3$ crystals to monoclinic by the structural transition at $T_s$ = 79 K. Upon further cooling another structural transition is observed at 32 K which causes lowering of the symmetry from monoclinic to triclinic. The transition at 79 K confirms previous reports [18,19] but contradicts the conclusions in Ref. 20. The structural transition at 32 K has not been reported so far. It is likely related to a transformation of the ferromagnetic structure, which has been recently reported by Gati et al. [21] at temperatures below 40 K. In contrast, the monoclinic symmetry remains conserved when $VI_3$ becomes ferromagnetic below $T_{FM1}$. Only a change of temperature dependence of lattice parameters is observed at temperatures below $T_{FM1}$ which is apparently a result of spontaneous magnetostriction. Our specific-heat measurements revealed that $T_s$ decreases in a magnetic field applied along the trigonal $c$-axis (by almost 3 K in 14 T) whereas it remains intact by the field perpendicular to $c$. This indicates that a magnetic field applied along the main axis of the $VI_6$ octahedron, assisted by magnetoelastic interactions, supports the hexagonal symmetry of the V honeycombs in the basal plane of VI₃. It is very likely that strong magnetoelastic interactions should be considered as the main underlying mechanisms driving the structure changes and transitions in VI$_3$.

The common structural motif of $TX_3$ compounds is a honeycomb net of $T$ cations as shown in Fig. S1 in SM [26]. The VI$_3$ RT trigonal structure of the BiI$_3$ structure is characterized by the ABC layer stacking sequence. The subsequent layers are shifted along one of the V-V "bonds". The honeycomb net is regular due to the three-fold symmetry.

Due to geometrical limitations of the LT diffraction setup and due to the fact that the sample has the shape of a thin (0 0 L) platelet, only diffraction maxima with L > 0 can be measured. Therefore, it is not possible to distinguish or determine the exact structure model of VI$_3$ by mapping selected diffraction peaks, but the temperature dependence of the lattice parameters can be studied.

In this study, we used the hexagonal unit cell for the trigonal crystal system and the corresponding H K L indices. The reciprocal space maps were measured around the symmetric (0 0 24), (0 0 21), (0 0 18), (0 0 15) and asymmetric (2 2 21), (2 2 18), (2 2 15), (1 1 21), (1 1 18), (1 1 15), (4 -2 21), (4 -2 18), (4 -2 15), (2 -1 21), (2 -1 15) diffraction peaks. The temperature dependence of the diffraction peaks is shown in Fig. 1. In the trigonal system, the (2 2 L) and (4 -2 L) diffraction peaks have the same $2\theta$, but different structure factor, so that the measured intensity ratio I$_{(2\ 2\ 21)}$/ I$_{(4\ -2\ 21)}$ ~10 is consistent with the trigonal crystal system. Below 79 K, the (2 2 L) and (4 -2 L) diffraction peaks are both split into two peaks, both split pairs having different $2\theta$ distances (Fig. 1,2). An additional diffraction-peak splitting is observed at temperatures below 32 K (Fig. 1, 2).

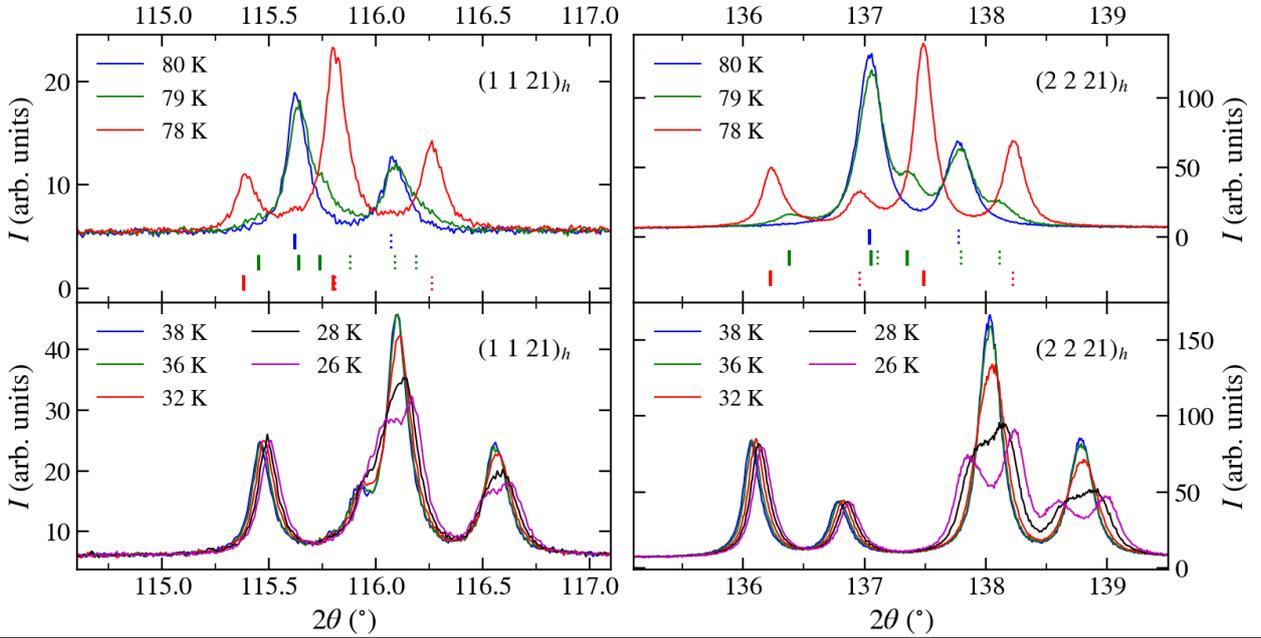

FIG. 1. Evolution of (1 1 21) (left panels) and (2 2 21) (right panels) diffraction peaks of a VI$_3$ single crystal in the temperature interval from 78 to 80 K (top) and from 26 to 38 K (bottom). The vertical solid and dotted markers indicate the Cu$_{K\alpha1}$ and Cu$_{K\alpha2}$ positions of the Cu$_{K\alpha1,2}$ doublet.

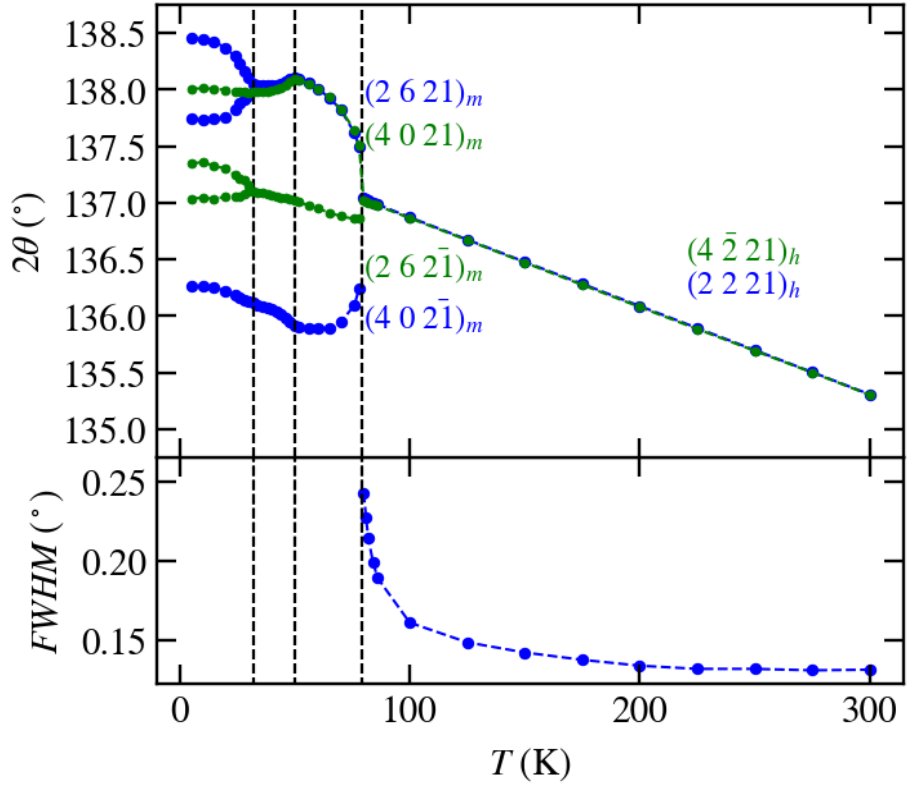

FIG. 2. Temperature dependence of the (2 2 21) and (4 -2 21) diffraction peaks (upper panel) and the diffraction-peak half-width *FWHM* of (2 2 21) (lower panel). The error bars are smaller than the markers.

The diffraction-peak splitting is ascribed to the reduction of the lattice symmetry. Such a reduction increases the number of non-equivalent diffraction peaks and the diffraction peaks belonging to the more symmetric structure disappear. Therefore, the appearance of the new diffraction peaks by a mere change of extinction rules cannot be explained. From Fig. 2, it is clear that with decreasing temperature one split diffraction peak moves to larger and the other to smaller $2\theta$ values. This fact can be understood by opposite deformation of non-equivalent domains.

The space group of the LT structure is expected to be a maximum "translationengleiche" (t-) subgroup of the trigonal crystal system. The trigonal space groups contain only monoclinic and triclinic t-subgroups. Two transitions have been observed and, therefore, it is assumed that the monoclinic lattice appears at higher temperatures of 32 – 79 K and the triclinic one below 32 K. We use the transformation formulas in (1) between the hexagonal and monoclinic unit cells, the indexes h and m stand for hexagonal and monoclinic unit cells, respectively.

$$\boldsymbol{a}_m = \boldsymbol{a}_h, \ \boldsymbol{b}_m = \boldsymbol{a}_h + 2\mathbf{b}_h, \ \boldsymbol{c}_m = \boldsymbol{c}_h. \qquad (1)$$

The hexagonal and monoclinic basis vectors are sketched in Fig. 3. It is worth to note that if $\beta = 90°$ and $b_m = \sqrt{3}\,a_m$, the same hexagonal primitive lattice is obtained.

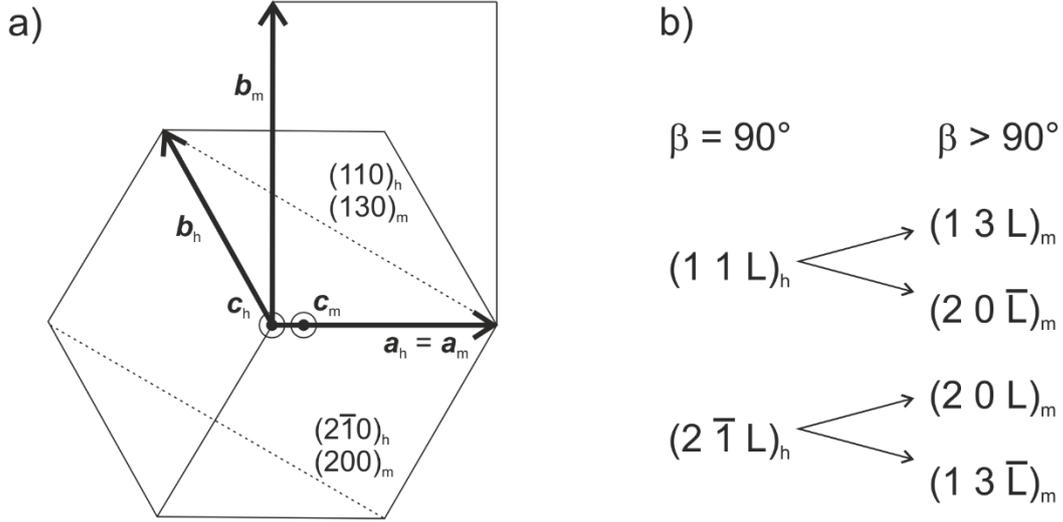

FIG. 3. a) Schematic representation, basal plane view, of HT and LT unit cell. The indices h and m label the hexagonal and monoclinic unit cell, respectively. b) Splitting of diffraction if $\beta > 90°$.

The lattice distortion is connected with the formation of domains in the sample. If we assume only $\beta \neq 90°$ and keep $b_m = \sqrt{3}\, a_m$ then the diffraction-peak $(2\ 2\ L)_h$ and $(4\ \text{-}2\ L)_h$ split into four diffraction peaks: $(2\ 2\ L)_h$, $(2\ 2\ \text{–}L)_h$, $(4\ \text{-}2\ L)_h$ and $(4\ \text{-}2\ \text{–}L)_h$, having different $2\theta$'s $((2\ 6\ L)_m, (2\ 6\ \text{–}L)_m, (4\ 0\ L)_m, (4\ 0\ \text{-}L)_m$ in the monoclinic system). This assumption is in agreement with observed split pairs $(2\ 6\ \text{–}L)_m, (4\ 0\ L)_m$ and $(4\ 0\ \text{-}L)_m, (2\ 6\ L)_m$ below 79 K, see Fig. 2. Fig. 3 also schematically shows, that if $\beta$ becomes different from 90° we will see diffraction peak $(1\ 3\ L)_m$ and from another domain the diffraction peak $(2\ 0\ \text{-}L)_m$ or $(1\ 3\ \text{-}L)_m$ and $(2\ 0\ \text{-}L)_m$. On the other hand, the symmetric diffraction peak $(0\ 0\ L)$ do not split, but the transitions are well visible in the change of the slope of the $2\theta(T)$ dependence (Fig. 4).

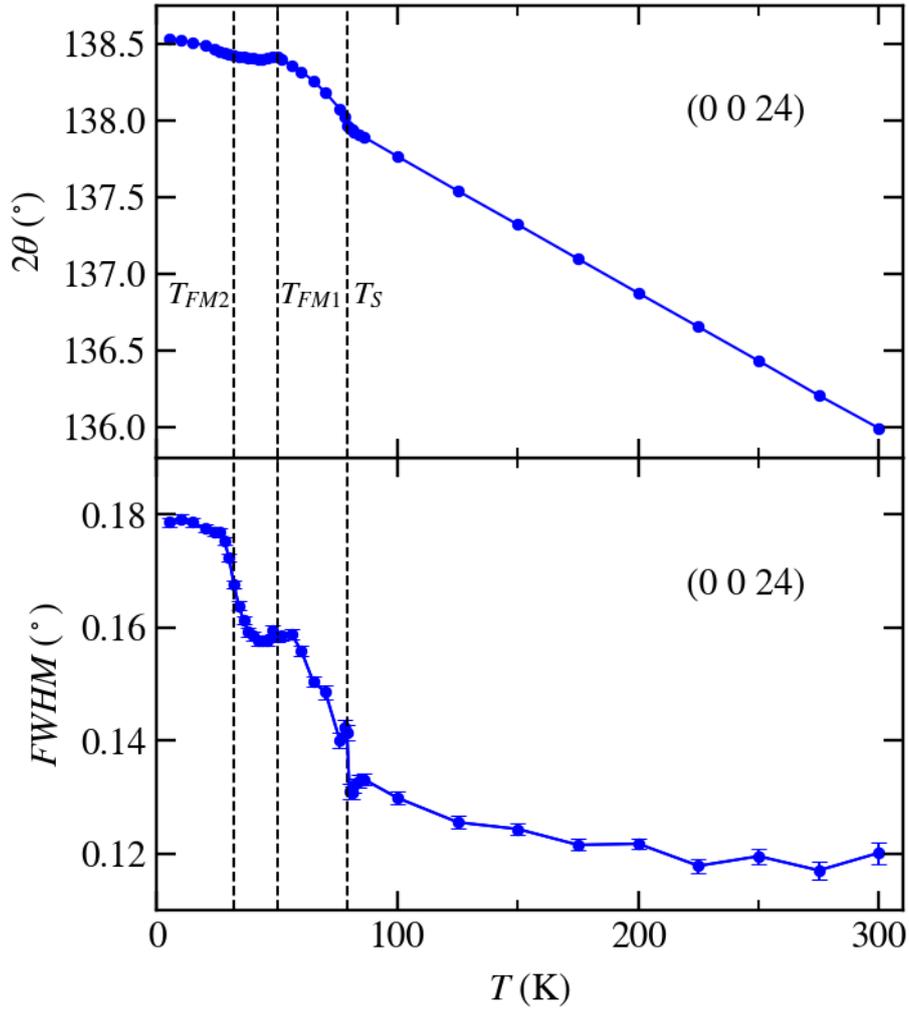

FIG. 4. Temperature dependence of the (0 0 24) diffraction peak (upper panel) and the diffraction - peak half-width (lower panel).

At RT, the mosaicity of the sample is around 2° (see the high-resolution reciprocal maps in the supplementary material) and therefore is better to integrate the maps in the rocking direction and use only $2\theta$ -curves for the determination of the lattice parameters. The correction for the sample displacement $\Delta 2\theta \sim \cos(\theta)$ has been used, which is strictly speaking only valid for symmetric diffraction peaks. Because of the simplicity, we used this correction factor also for asymmetric diffraction peaks. This results in a small systematic error of the lattice parameters, but their temperature dependence remains unaffected. The resulting temperature dependences of the lattice parameters are shown in Fig. 5. Comparison of measured data and fits is presented in the SM [26].

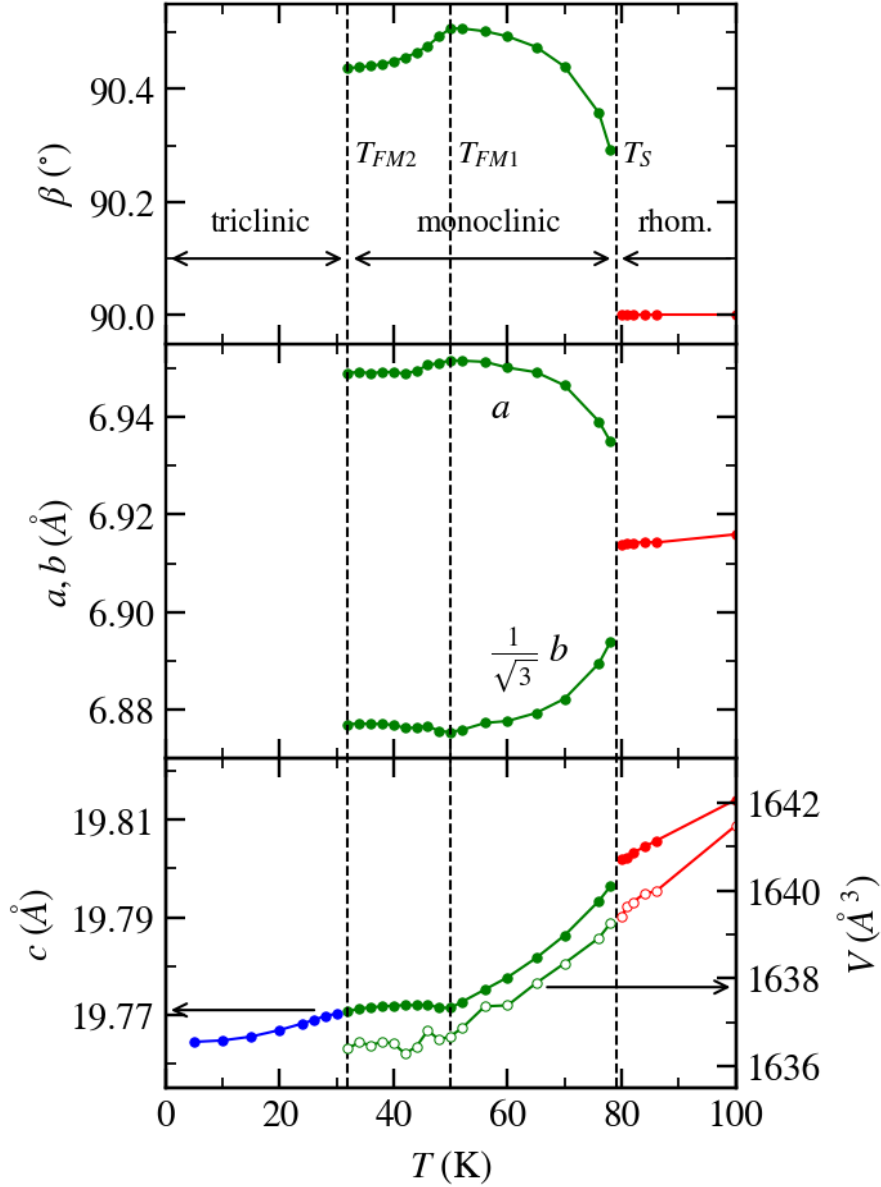

FIG. 5. Temperature dependence of lattice parameters and unit-cell volume of VI$_3$.

In Fig. 6, a comparison of the evolution of the specific heat, magnetization and two diffraction peaks is shown. One can recognize a dramatic splitting of the diffraction-peaks coinciding with a sharp specific-heat anomaly at $T_s$ whereas only negligible change of magnetization was observed at this temperature. The specific-heat and magnetization anomalies at $T_{FM1}$ are well pronounced; no splitting of diffraction peaks is seen, but the character of the temperature dependence of the diffraction angles is changed. Clear additional splitting of certain diffraction peaks, which indicates lowering of the crystal symmetry from monoclinic to triclinic, is observed below a temperature, which we denote as $T_{FM2}$ (= 32 K). Only slight changes in temperature dependence of the specific heat and magnetization are observed around this temperature.

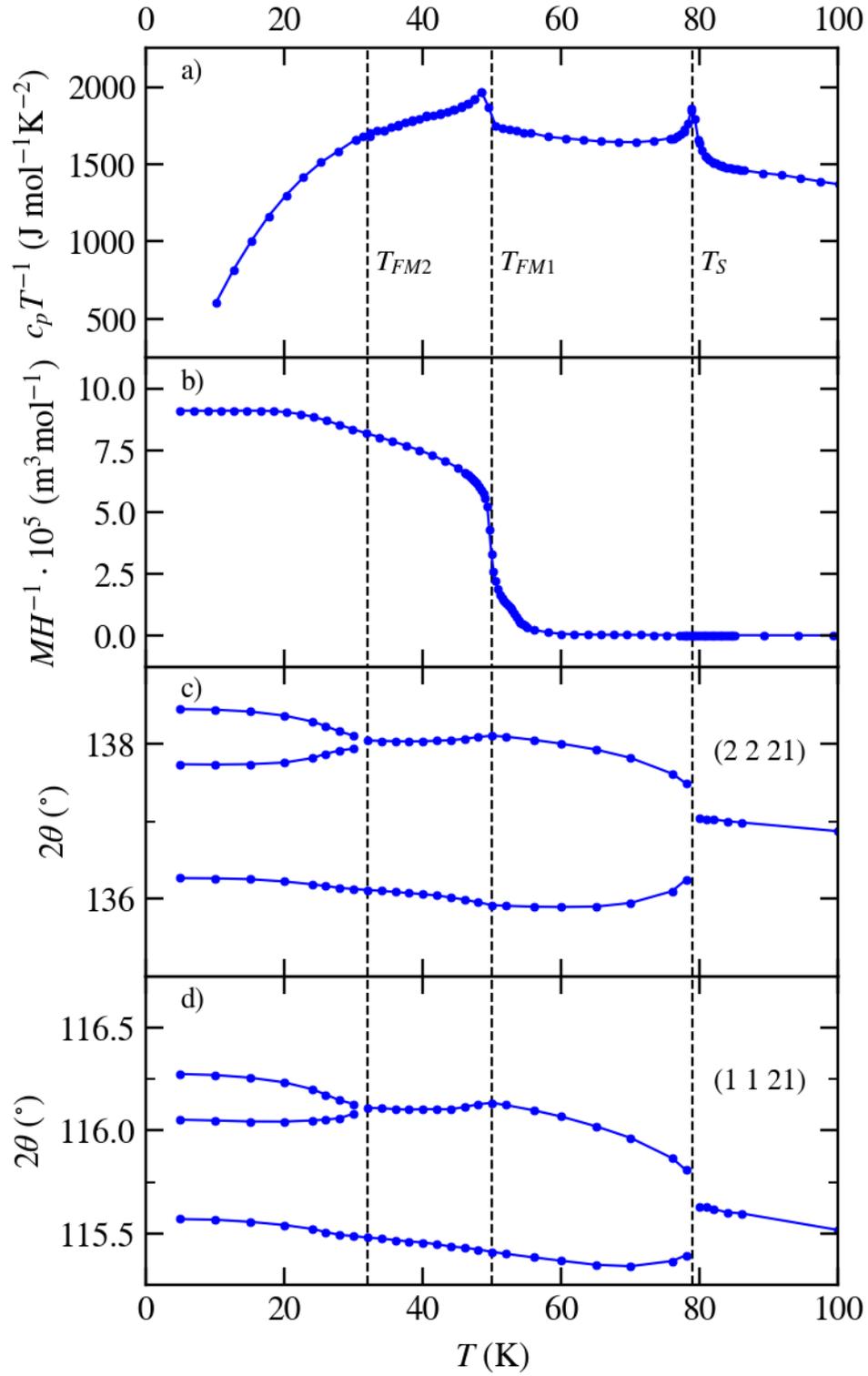

FIG. 6. Temperature dependence of a) specific heat in a $C_p/T$ vs $T$ plot, b) c-axis magnetic susceptibility in a magnetic field of 1 mT, c) (2 2 21) and d) (1 1 21) diffraction peak of a VI$_3$ crystal.

The decrease of $T_s$ with increasing magnetic field shown in the inset in Fig. 7 seems to be accelerated in fields above 7 T. This is most probably reflecting enhanced polarization of V paramagnetic moments in high fields. A possibility that it is due to the $C_p$ anomaly associated with the FM ordering is skewed up in temperature to reach $T_s$ requires more detailed studies of specific heat in magnetic fields.

No field dependence of the anomaly is observed in a field applied within the *ab*-plane as can be seen in Fig. S15 in Supplemental material [26].

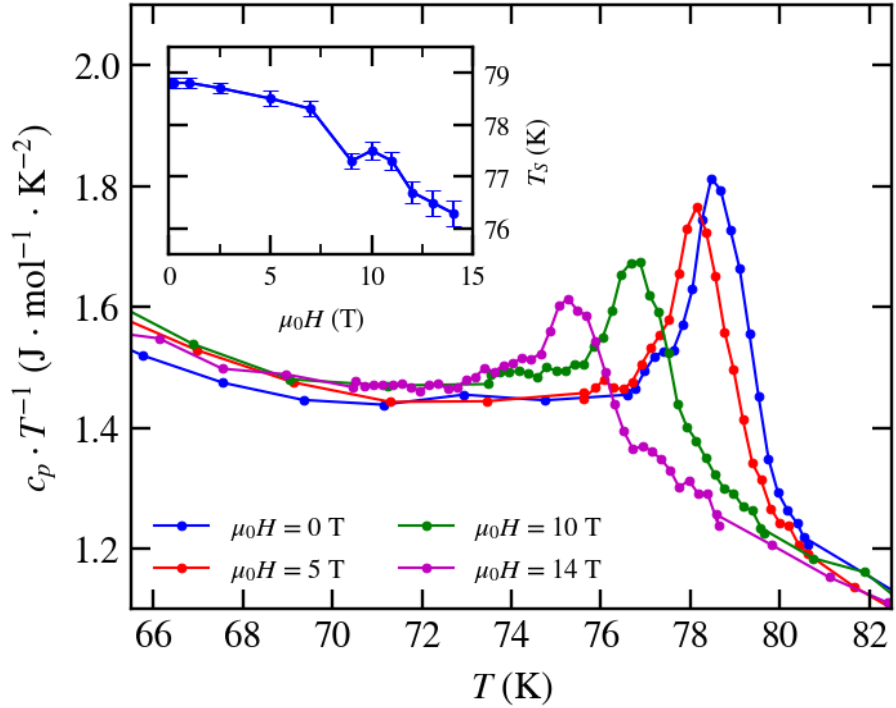

FIG. 7. Temperature dependence of the specific heat ($C_p/T$ vs. $T$ plot) of VI$_3$ in the vicinity of $T_s$, measured in various magnetic fields applied parallel to the *c*-axis. Inset: Magnetic-field dependence of $T_s$.

The RT structure of VI$_3$, refined from the XRSCD data, is trigonal which is in agreement with most literature sources [14,15,18,19]. The same *R-3* (148) space group has been reported in [13,15,19]. The trigonal structure persists upon cooling down to 80 K as evidenced by the unchanged set of measured diffraction peaks. As seen in Fig. 6 and relevant figures in SI, the temperature dependences of the diffractions angles between RT and 80 K are almost linear.

The sudden splitting of the (1 1 L)$_h$ and (2 -1 L)$_h$ diffraction peaks observed upon cooling the VI$_3$ crystal from 79 to 78 K (see Fig. 6 and relevant figures in SM [26]) is a result of a structural transition characterized by a lowering of the crystal symmetry, specifically the HT trigonal structure transforms below 80 K to a lower-symmetry structure which can be described by a monoclinic lattice. The transition is the first order phase transition, because we have observed both, the rhombohedral and the monoclinic phase, coexisting at 79 K (Fig. 1 top). This coexistence of phases is not compatible with the second order phase transition. Also the position of diffraction peaks corresponding to rhombohedral and monoclinic phase are well separated (Fig. 1 top). The transition itself looks like a disappearing of rhombohedral peak and appearing of monoclinic peaks at different $2\theta$. On the other hand, the temperature cycles around the transition do not show any visible hysteresis (FIG. S16).The lattice parameters $a,b,c$ and $\beta$ change abruptly at $T_s$ but the lattice volume remains unchanged.

This structural transition at $T_s = 78\text{-}79$ K has been already reported in recent papers [18-219] as accompanied by a sharp specific-heat anomaly and a small negative step in the $c$-axis magnetization. The measurements of the temperature dependence of the specific heat in the applied magnetic field reveal surprising behavior. $T_s$ is found to be dependent on a magnetic field applied along the $c$-axis, in particular it is decreased by almost 3 K in 14 T. On the other hand, a field applied in perpendicular direction has a negligible effect on the $T_s$ related anomaly. We presume that this behavior is of magnetoelastic origin and reflects that the magnetic field applied along the axis of the $VI_6$ octahedra, assisted by magnetoelastic interactions, is protecting the hexagonal symmetry of the honeycomb network in the $VI_3$ lattice which leads to stabilization of the trigonal structure to somewhat lower temperatures.

A transition to the ferromagnetic state is usually accompanied by a subtle distortion of the crystal lattice in zero magnetic field. This spontaneous magnetostriction effect is due to magnetoelastic interactions. This is presumably also the case in $VI_3$. In Fig. 5 one can see a slight increase of $b$ and c and decrease $a$ and $\beta$, which increases below $T_{FM1}$ towards to 90 ° of the monoclinic structure. The volume magnetostriction between $T_{FM1}$ and 32 K is slightly positive, which is frequently observed in ferromagnets. The distortions induced below $T_{FM1}$ by spontaneous-magnetostriction frequently cause symmetry breaking. There are, however, not indications for this in the present study.

However, upon cooling across 32 K, further lowering of the $VI_3$ crystal symmetry, namely from monoclinic to triclinic, is observed. The monoclinic-triclinic transition at $T_{FM2} \sim 32$ K has not been reported so far. We associate this transition of the $VI_3$ crystal structure with the transformation of the ferromagnetic structure below 36 K, which has been recently reported by Gati et al. [21] from a thorough NMR study. The connections of structural transitions with magnetic phenomena in $VI_3$ suggest a considerable role of magnetoelastic interactions in this compound. This idea can be further specified when knowing microscopic aspects of the magnetic structures which are most probably not simple. Somewhat canted ferromagnetic structures have been suggested in previous papers [19,21]. Therefore, investigations focused on details of the magnetism of $VI_3$ are strongly desired.

In conclusion, the present extended XRSCD study of the crystal structures and structural phase transitions of the van der Waals ferromagnet $VI_3$ confirms the existence in three temperature regions and their evolution around phase transitions the existence of the trigonal structure of the $R\text{-}3$ (148) space group at 250 K. Upon cooling, the sudden splitting of certain diffraction peaks at temperatures between 79 and 78 K unambiguously confirms that $VI_3$ undergoes structure phase transition between the HT trigonal and the LT monoclinic phase at $T_s \sim 79$ K. The critical temperature of this transition $T_s$ has been found to decrease in magnetic fields applied along the trigonal $c$-axis. Considerable magnetostriction-induced changes of the monoclinic-structure parameters have been recorded below 50 K (= $T_{FM1}$) when $VI_3$ becomes ferromagnetic. The monoclinic structure transforms upon further cooling into a triclinic variant at 32 K. This transition most probably occurs in conjunction with a transformation of the ferromagnetic structure. These phenomena are associated with a strong magnetoelastic coupling in $VI_3$. For further understanding of the complex structure and the magnetic phenomena in $VI_3$, further experiments at neutron and synchrotron infrastructures focused on microscopic aspects of the magnetism are highly desired.

The authors would like to thank Prof. de Boer for critical reading and correcting the manuscript and Dr. Petříček for valuable discussions about the peculiarities of trigonal symmetry. This work is a part of the bilateral Czech – Korean research project which is financed by the Czech Science Foundation grant GACR 19-16389J and by the grant IBS-R009-G1 provided by the Institute for Basic Science of the Republic of Korea. Most of the experiments were carried out in the Materials Growth and Measurement Laboratory MGML (see: http://mgml.eu) which is supported within the program of Czech Research Infrastructures (project no. LM2018096).The single-crystal analysis was supported by the project 18-10438S of the Czech Science Foundation using instruments of the

Supplemental material for:

# Crystal Structures and Phase Transitions of the van-der-Waals Ferromagnet VI₃

P. Doležal[1], M. Kratochvílová[1], V. Holý[1], P. Čermák[1], V. Sechovský[1], M. Dušek[2], M. Míšek[2], T.Chakraborty[3],Y. Noda[4], Suhan Son[3,5], Je-Geun Park[3,5]

[1]*Charles University, Faculty of Mathematics and Physics, Department of Condensed Matter Physics, Ke Karlovu 5, 121 16 Prague 2, Czech Republic*
[2]*Institute of Physics, Academy of Sciences of Czech Republic, v.v.i, Na Slovance 2, 182 21 Prague 8, Czech Republic*
[3]*Center for Correlated Electron Systems, Institute for Basic Science, Seoul 08826, Korea*
[4]*Institute of Multidisciplinary Research for Advanced Materials, Tohoku University, Sendai 980-8577, Japan*
[5]*Department of Physics and Astronomy, Seoul National University, Seoul 08826, Korea*

*Experimental methods used*

The single crystals of VI₃ were prepared by the chemical vapor transport method, as described elsewhere [1].

The room-temperature phase was determined by XRSCD study on a VI₃ single crystal using an X-ray diffractometer SuperNova, equipped with a Mo X-ray micro-focus tube with a mirror monochromator producing Mo Kα radiation, and a CCD detector Atlas S2. Due to the instability of the sample at room temperature, data were collected at 250 K on a sample covered by a protective oil (perfluoropolyalkylether). Experimental details are given in Table 1. We used programs CrysAlis [2] for data collection and processing, Superflip for structure solution [3] and Jana2006 [4] for structure refinement. Structure analysis was complicated because of the layered nature of the sample, which leads to large mosaicity causing smearing of the diffraction peaks along c*. Nevertheless, after testing many samples, we were able to find a specimen with good diffraction patterns was found. All tested samples exhibited twinning.

We also measured selected symmetric high-resolution (HR) ω-2θ scans and reciprocal-space maps at RT using a RIGAKU SMARTLAB 9KW rotating-anode diffractometer and a PaNalytical MRD sealed-tube diffractometer. Both systems were equipped with a 2×220Ge channel-cut monochromator and a two-dimensional detector. These measurements were focused on the crystal quality (mosaicity, random fluctuations of the widths of the van der Waals gap. In order to prevent possible deterioration on the sample structure by long exposure to air, the sample was covered by a thin kapton foil during the diffraction measurements.

The temperature dependence of the lattice parameters from 5 to 300 K was determined using the reciprocal space maps of selected diffractions of the single-crystalline sample. The reciprocal maps were measured in a refurbished Siemens D500 θ − θ diffractometer equipped with a Mython 1K position-sensitive detector in Bragg-Brentano geometry using the non-monochromatic Cu$_{Kα1,2}$ radiation. The single-crystalline sample was placed on the piezo-rotator (which allows aligning the sample in ϕ direction) connected to the „cold finger" of the cryostat (*ColdEdge*). To reach good thermal stabilization, the inner cap was filled with He gas. As a cooling source a Gifford-McMahon refrigerator was used. The stabilization of temperature was better than 0.1 K with an absolute uncertainty of 0.5 K and with the lowest possible temperature of 3 K.

The construction of the low-temperature diffractometer is suitable for X-ray powder diffraction (XRPD) which does not provide entire crystal structure refinement. The VI$_3$ crystals were available in the form of very thin and soft plates with the trigonal $c$-axis perpendicular to the plate. The mechanical properties prevent the preparation of a reasonable powdered sample. Pulverization would cause major damage to the structure. Therefore, we have decided to focus on the single-crystalline samples as they were and measured the reciprocal space maps of selected diffractions. The only possible alignment of the sample was with the $c$-axis in the scattering plane. Therefore, direct measurement of the lattice parameter $a$ by the mapping of (H 0 0) diffractions is not possible.

The temperature dependence of the magnetization was measured between 2 and 100 K using an MPMS-7 apparatus (*Quantum Design, Inc.*). The specific-heat data were collected in the temperature range 2-300 K in magnetic fields up to 14 T using PPMS-14 (*Quantum Design, Inc.*).

*Room-temperature crystal structure of VI$_3$*

The common structural motif of *TX$_3$* compounds is a honeycomb net of T cations that are at edges sharing octahedral coordination, as shown in Fig. S1. At room temperature, VI$_3$ has the trigonal structure of the BiI$_3$ structure with layer-stacking sequence ABC. The subsequent layers are shifted along one of the V-V "bonds". The honeycomb net is regular due to the three-fold symmetry.

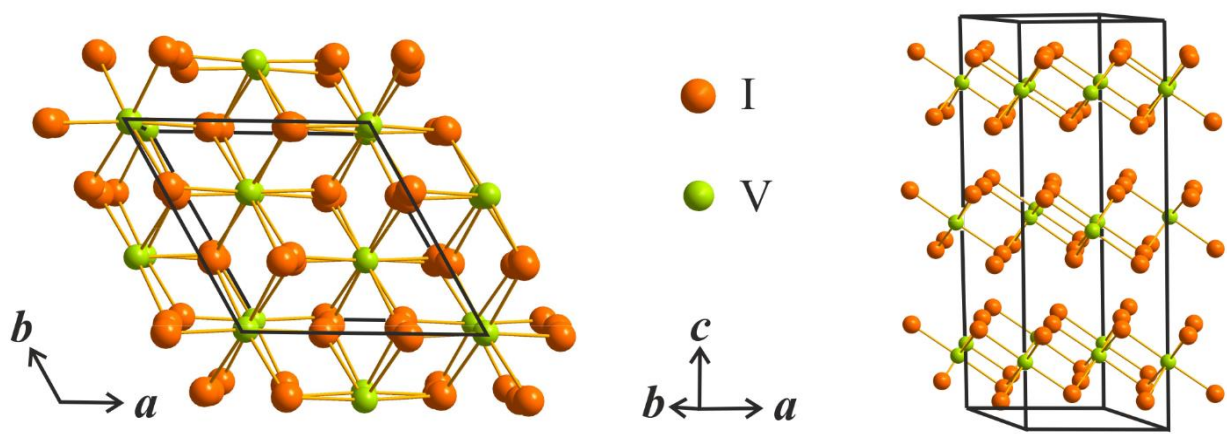

Fig. S1. Schematic picture of the room-temperature rhombohedral structure of VI$_3$

*Refinement of the room-temperature structure*

The symmetry test of Jana2006 calculated the following R$_{int}$ values for Laue symmetries compatible with the cell parameters: R$_{int}$ ~ 0.3 for 6/mmm, 6/m and -3m1, R$_{int}$ ~ 0.08 for -31m, and R$_{int}$ ~ 0.07 for -3. Thus, the possible Laue symmetries were -31m or -3, where the former cannot be combined with R centering, which is only possible for -3m1 (excluded by the high R$_{int}$), and -3. The conditions for systematic absences due to R centering were violated by ~2000 of relatively weak diffraction peaks with average I/σ(I) about 7 for both obverse and reverse setting. The crucial question was, therefore, whether the violation of the R centering occurs due to twinning or whether the R centering is really violated in the crystal structure.

For Laue symmetry -3, four-fold twinning is possible as a result of symmetry lowering from the highest 6/mmm. The twinning matrices can be formulated as follows

$$T1 = \begin{pmatrix} -1 & 0 & 0 \\ 0 & -1 & 0 \\ 0 & 0 & 1 \end{pmatrix}, \quad T2 = \begin{pmatrix} 1 & 0 & 0 \\ -1 & -1 & 0 \\ 0 & 0 & -1 \end{pmatrix}, \quad T3 = \begin{pmatrix} -1 & 0 & 0 \\ 1 & 1 & 0 \\ 0 & 0 & -1 \end{pmatrix},$$

where T1, T2 and T3 represent a two-fold axis along (0,0,1), (1,0,0) and (1,2,0), respectively. The symmetry test based on the -3 Laue symmetry and taking into the account these twinning operations has dramatically reduced the number of diffraction peaks violating the R centering (from ~2000 to ~250) and their average $I/\sigma(I)$ (from ~7 to ~5.5). This confirms the Laue symmetry -3 and discards the Laue symmetry -31m, which cannot be combined with R centering.

Refinement of the structure with the space group R-3 quickly converged with excellent R values (see Table S1) and the refined twin-volume fractions 0.0085(6), 0.0076(6) and 0.3921(17) for T1, T2 and T3, respectively. It should be noted, that the commonly observed reverse twinning by a two-fold rotation along (0,0,1) is described by the twinning matrix T2, with almost zero corresponding twin-volume fraction. The dominant twinning described by T3 seems to us rather unusual in comparison with often occurring twinning by T1. Another interesting point is that the weakly occupied twin volumes nevertheless decrease the R-value by 0.004 (from ~0.03 to 0.026) so that they are probably also present in the specimen. Other investigated samples exhibited the same kind of twinning but with different twin-volume fractions.

In the light that Reference [5] reports crystal structures of this material based on the Laue symmetry -31m, refined successfully from powder data, we have tried to ignore the fact of the R centering and determined the structure using space group P-31m. Although the refinement was very unstable and ADP parameters of iodine atoms mostly wrong, the resulting R-value was about 0.06 This demonstrates that even an incorrect structure can provide quite a good fit in Rietveld refinement. In this case, the accuracy is considerably lower compared with structure determination from single-crystal diffraction data.

Table S1

| Crystal data | |
|---|---|
| Chemical formula | $VI_3$ |
| $M_r$ | 431.7 |
| Crystal system, space group | Trigonal, *R* |
| Temperature (K) | 250 |
| $a, c$ (Å) | 6.9257 (3), 19.9185 (13) |
| $V$ (Å$^3$) | 827.40 (7) |
| $Z$ | 6 |
| Radiation type | Mo $K\alpha$ |
| $\mu$ (mm$^{-1}$) | 18.41 |
| Crystal size (mm) | $0.23 \times 0.12 \times 0.02$ |
| | |
| **Data collection** | |
| Diffractometer | SuperNova, Dual, Cu at home/near, AtlasS2 |
| Absorption correction | Numerical absorption correction based on gaussian integration over a multifaceted crystal model, combined with an empirical absorption correction using spherical harmonics. |
| $T_{min}, T_{max}$ | 0.15, 0.918 |
| No. of measured, independent and | 6954, 792, 688 |
| observed [$I > 3\sigma(I)$] reflections | |
| $R_{int}$ | 0.058 |
| $(\sin\theta/\lambda)_{max}$ (Å$^{-1}$) | 0.684 |
| | |
| **Refinement** | |
| $R[F^2 > 3\sigma(F^2)]$, $wR(F^2)$, $S$ | 0.026, 0.061, 1.21 |
| No. of reflections | 792 |
| No. of parameters | 16 |
| $\Delta\rho_{max}$, $\Delta\rho_{min}$ (e Å$^{-3}$) | 1.02, −0.74 |

*High-resolution X-ray diffraction*

The mosaicity of the sample was investigated by measuring high-resolution reciprocal-space maps of diffracted intensity around two chosen asymmetric reciprocal lattice points (Fig. 2). In reciprocal space, the diffraction maxima have the form of narrow arcs (parts of the Debye rings); the angular width of the maximum along the ring equals the angular mosaicity of the lattice. The mosaicity angle of 2° is denoted by the blue dashed radial lines. From the fact that the diffraction maxima are quite narrow in the radial direction (their width is comparable to the device resolution) we conclude that the relative fluctuation of the lattice parameters is smaller than 10$^{-3}$.

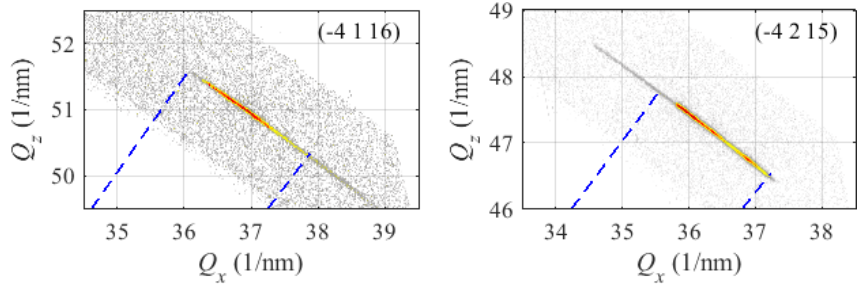

FIG. S2. High-resolution reciprocal-space maps of the diffracted intensity measured in the vicinity of two reciprocal-lattice points (in brackets). The colors span over four decades. The blue dashed lines denote the radial directions towards the origin of reciprocal space.

We have compared the measured symmetric HR $\omega$-$2\theta$ with simulations, assuming that the widths of the van der Waals gaps randomly fluctuate with a given mean value of 3.4446 Å and root-mean square deviation $\sigma$; Fig. 3 shows the results. In the simulations we took into account the actual resolution function of the diffractometer and we used the calculation routine based on kinematical-diffraction approximation (including refraction and absorption) and a polycrystal-like description of a random sequence of basal (001) atomic planes. The method is described in detail in Ref. [6]. The free (adjusted) parameters were the rms deviation $\sigma$, the thermal B-factor, and the primary intensity; the best match could be achieved for $\sigma = 0.03$ Å. Such small fluctuations affect not only the heights of the symmetric diffraction maxima 00L, but also the broadening of the basis of the peaks. The broad maximum at approx. $2\theta = 20°$ is caused by the kapton foil.

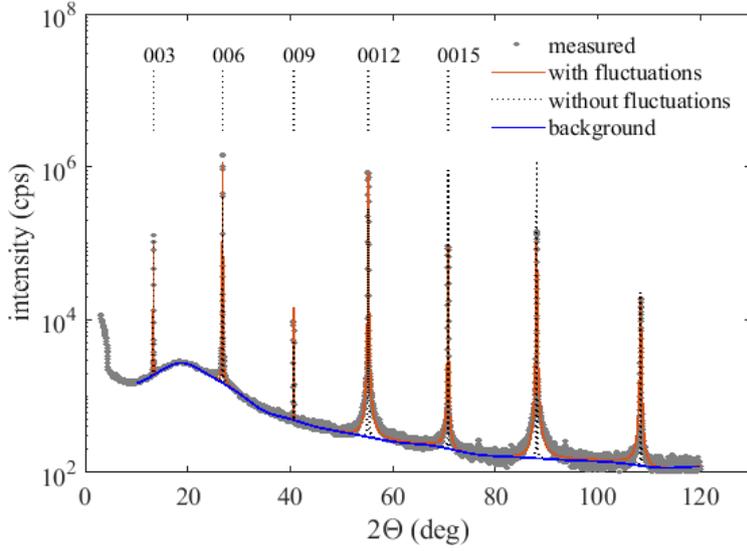

FIG. S3. Measured (points) and simulated (lines) HR $\omega$-$2\theta$ symmetric scans.

Finally, we inspected the long-term stability of the lattice during air exposure. For this purpose, we measured integrated intensities of selected symmetric $00l$ diffraction maxima on an un-covered sample at room temperature; the results are plotted in Fig. 4. The intensities roughly decay exponentially as $\exp(-\frac{t}{\tau})$ with the time constants $\tau$ between 15 and 20 hours. The microscopic mechanism of the lattice deterioration is not clear; the decrease of the integrated intensity might be ascribed to a gradual deterioration of the surface layers, which would decrease the diffracting sample volume. In this case however, all relative intensities would follow the same time dependence. The intensity decrease may also be explained by a progressive increase of the lattice disorder. This effect would result in the decrease of the static Debye-Waller factor $DW = \exp\left[-\frac{1}{2}(\sigma Q)^2\right]$ and the time constant $\tau$ would be inversely proportional to the square of the third diffraction index $L$.

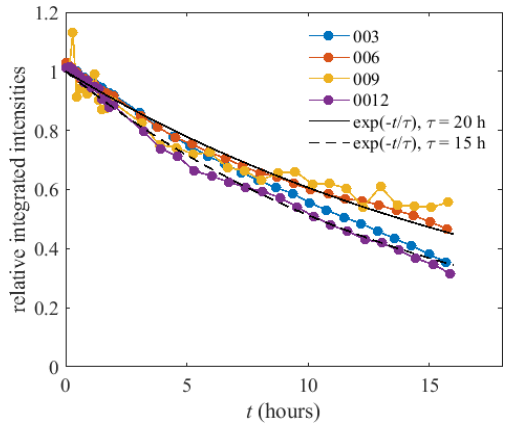

FIG. S4. Time dependence of the integrated intensities of various symmetric (0 0 L) diffraction maxima. The intensities are normalized to their initial values. Two exponential-decay curves are plotted by full and broken black lines.

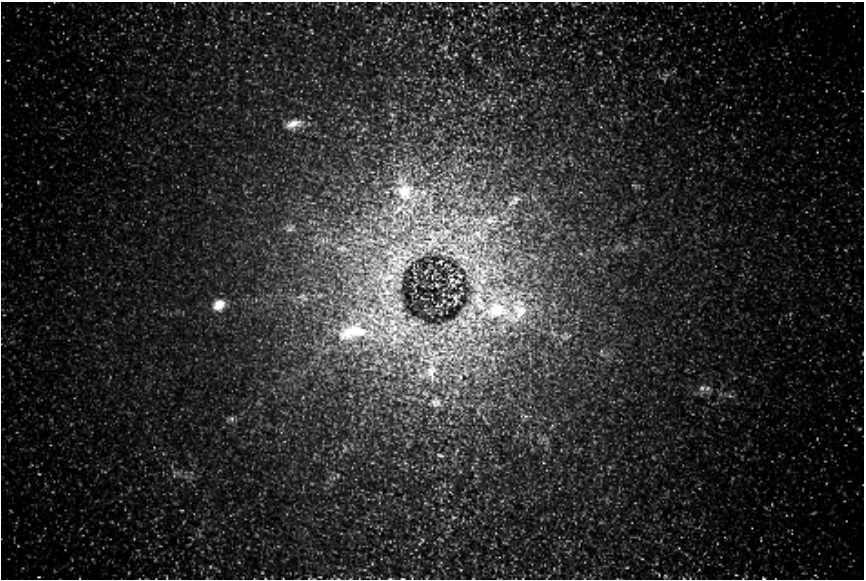

FIG. S5. Room-temperature X-ray Laue pattern of a VI$_3$ single crystal

*Low-temperature X-ray diffraction*

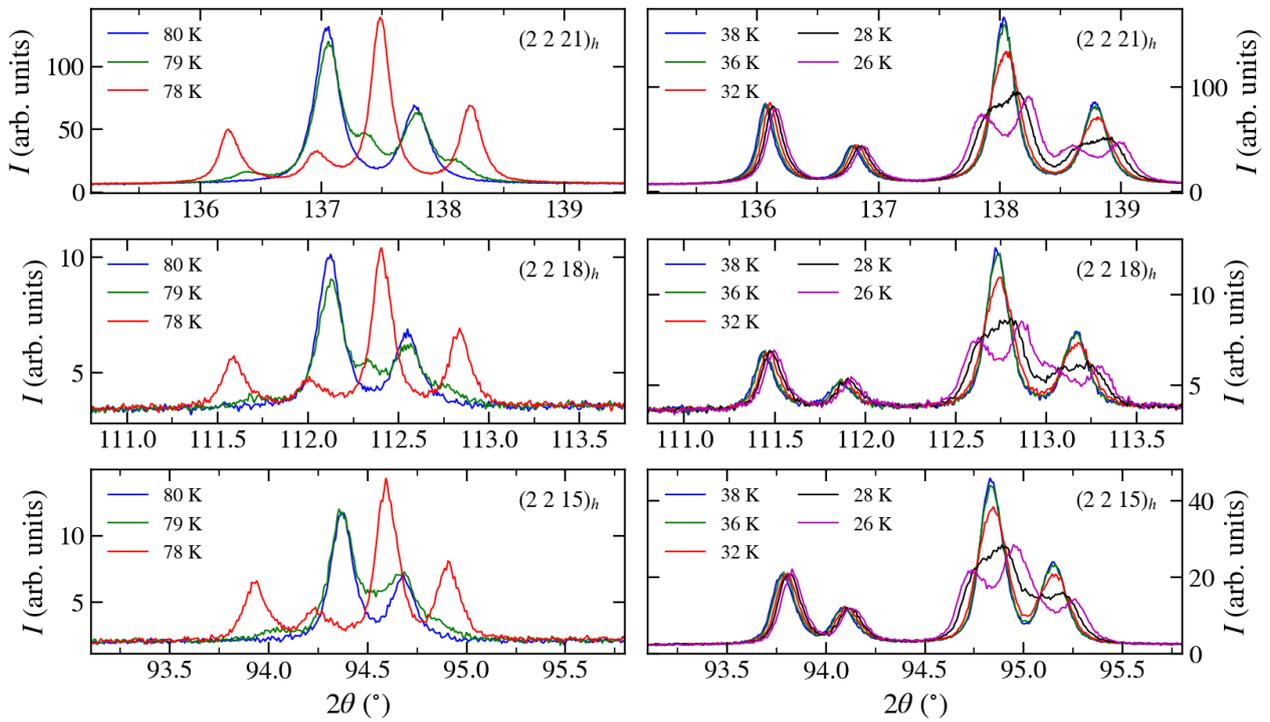

FIG. S6. Evolution of (2 2 L) (L = 21, 18, 15) diffraction peaks of the VI$_3$ single crystal in the temperature interval from 78 to 80 K (left panel) and from 26 to 38 K (right panel).

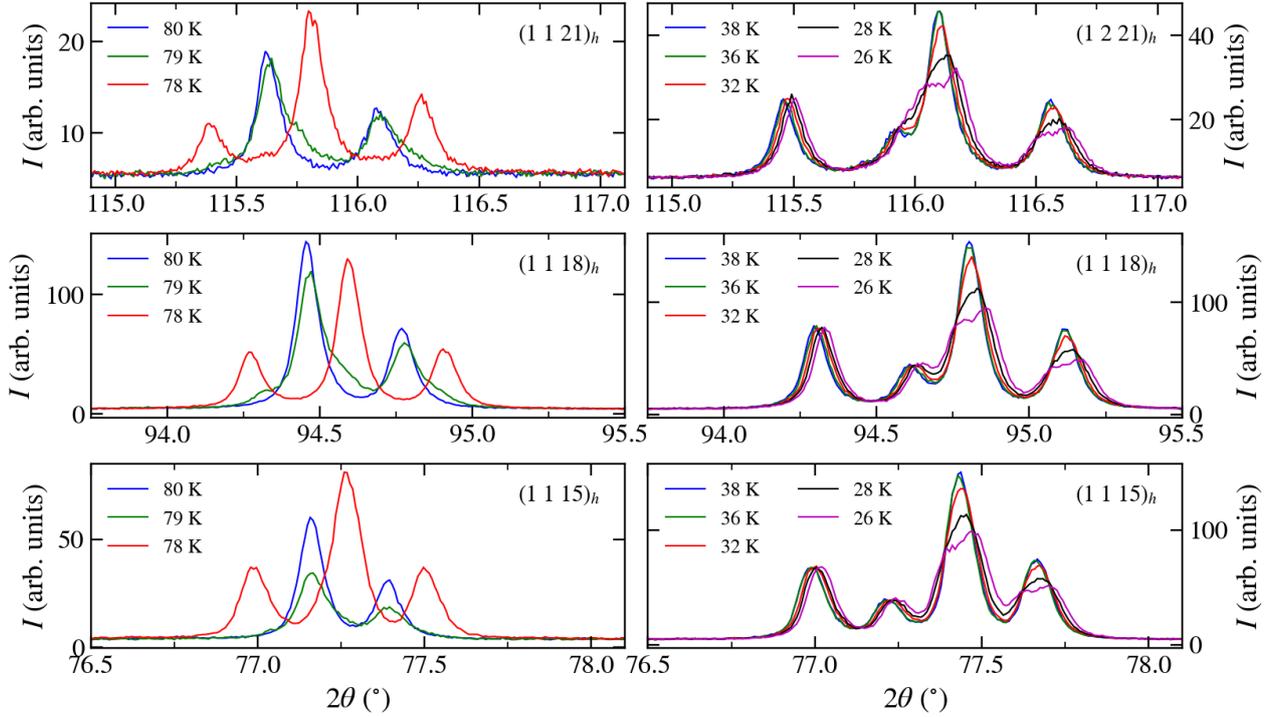

FIG. S7. Evolution of the (1 1 L) (L = 21, 18, 15) diffraction peaks of the VI$_3$ single crystal in the temperature interval from 78 to 80 K (left panel) and from 26 to 38 K (right panel).

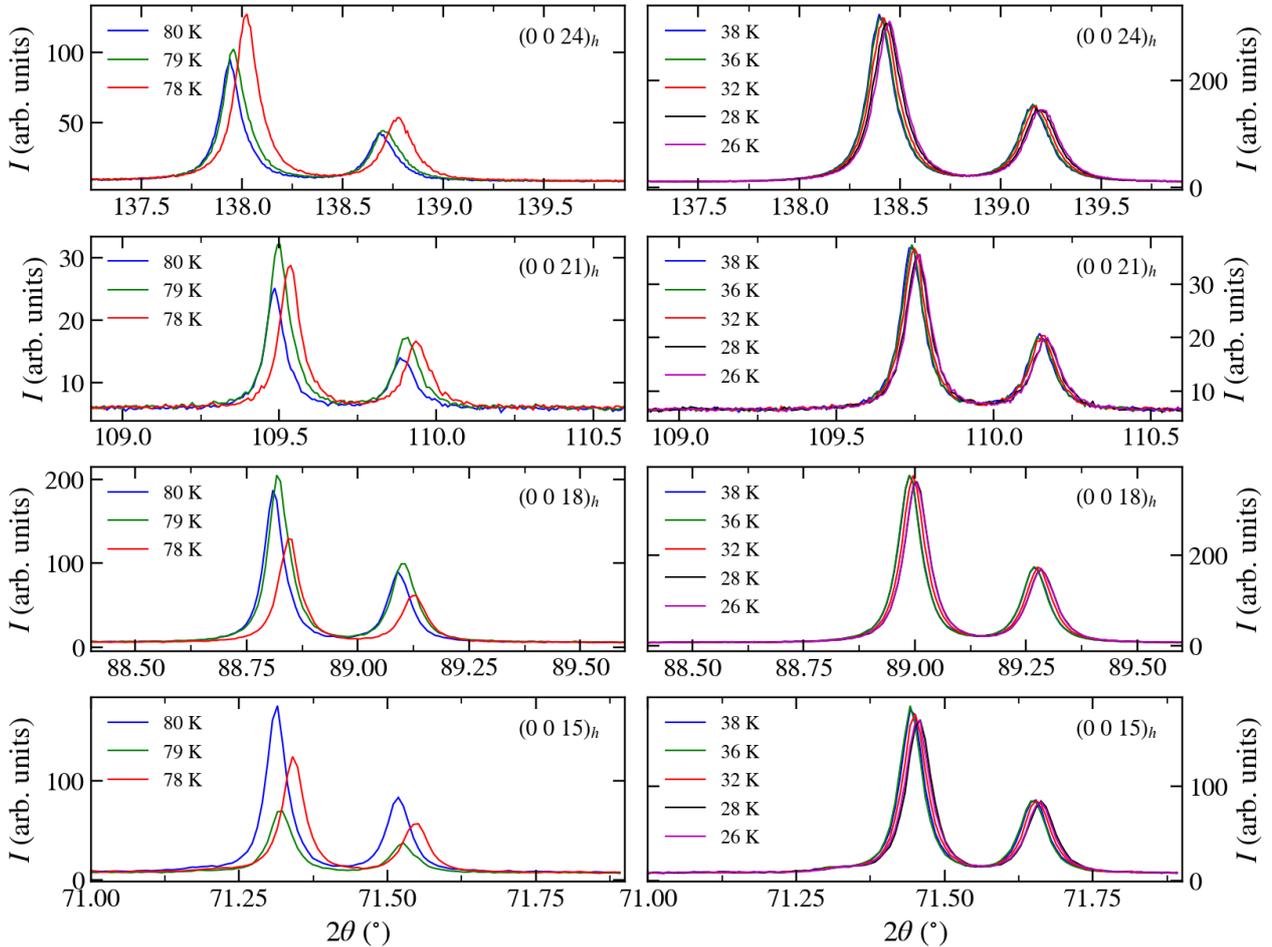

FIG. S8. Evolution of (0 0 L) (L = 24, 21, 18, 15) diffraction peaks of the VI₃ single crystal in the temperature interval from 78 to 80 K (left panel) and from 26 to 38 K (right panel).

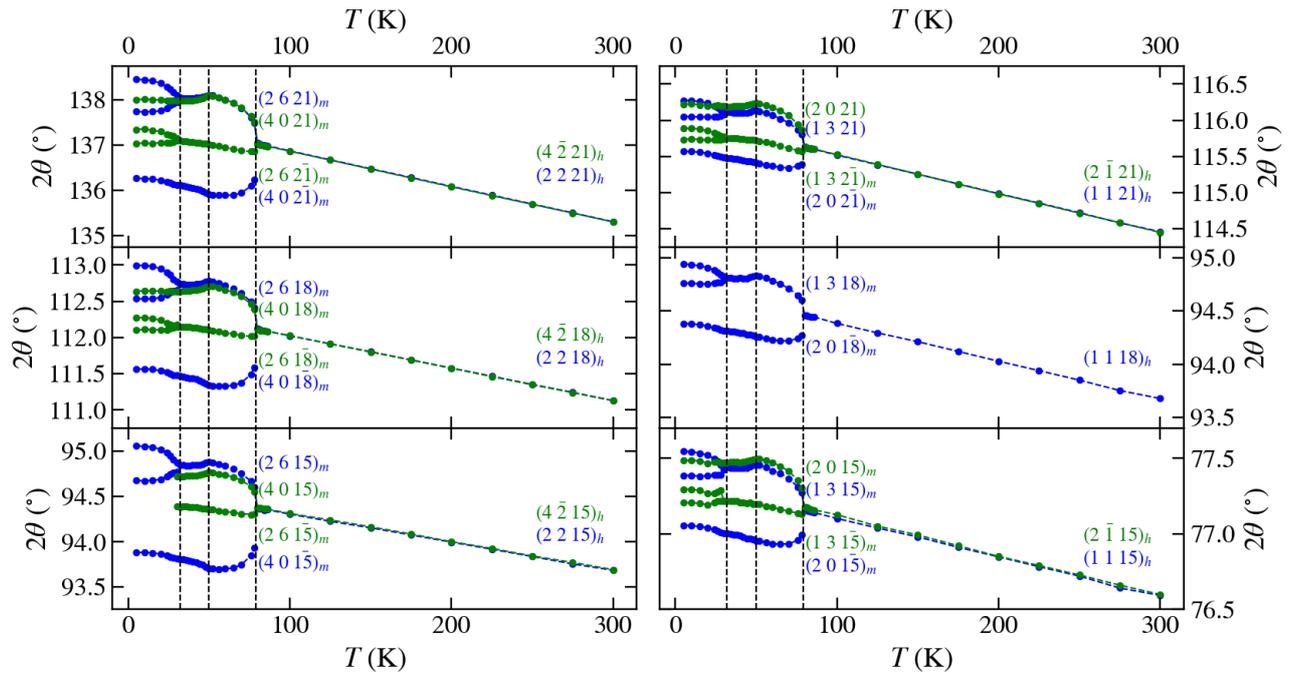

FIG. S9. Temperature dependence of the (2 2 L) and (1 1 L) (L = 21, 18, 15) diffraction peaks.

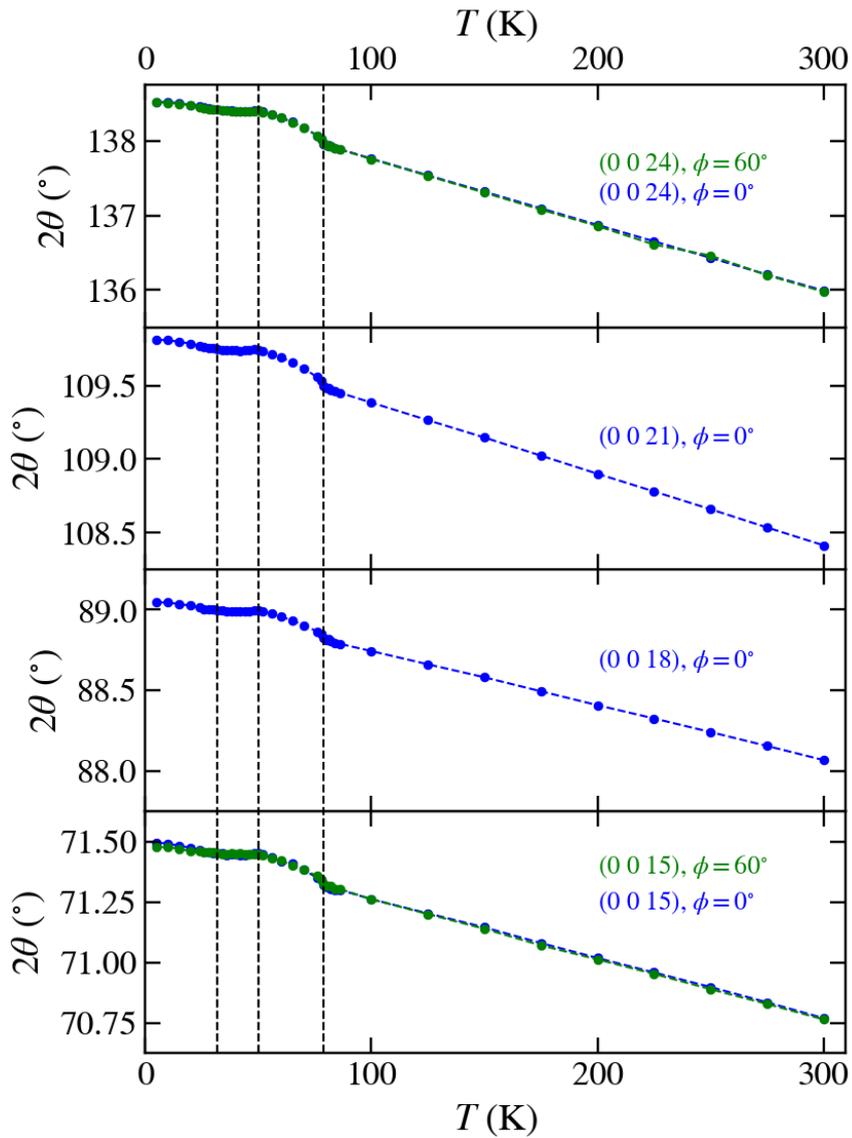

FIG. S10. Temperature dependence of the b(0 0 L) (L = 24, 21, 18, 15) diffraction peaks.

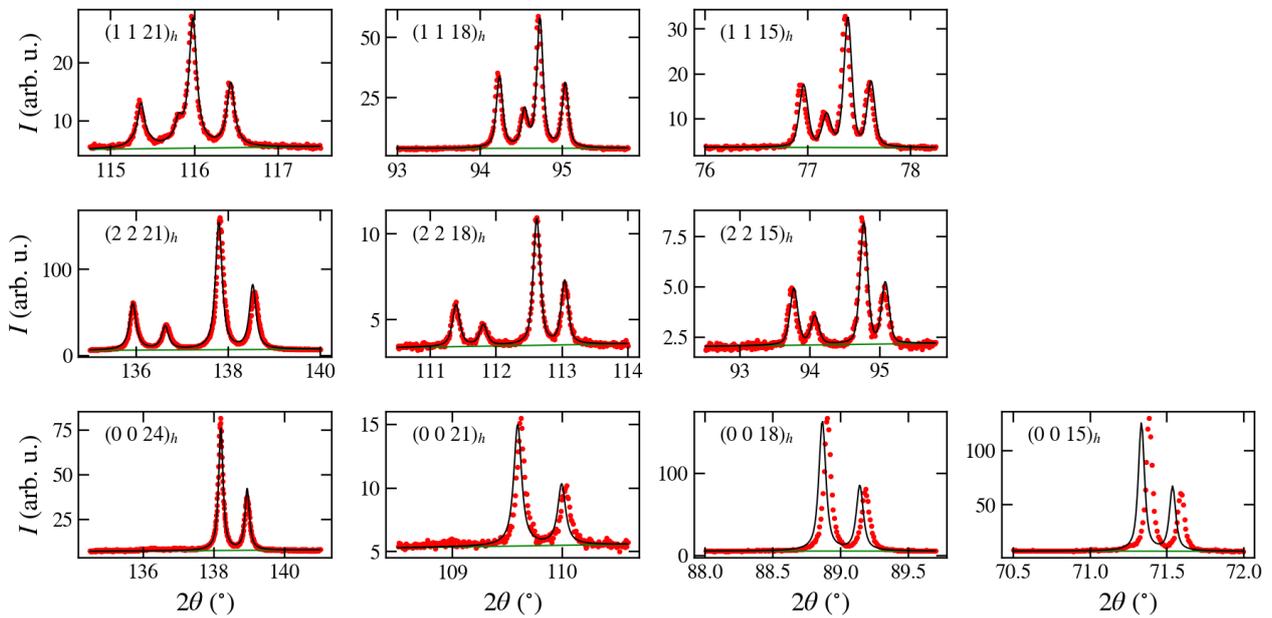

FIG. S11. The comparison of measured data and fits for (1 1 L), (2 2 L) and (0 0 L) (L = 21, 18, 15) diffractions at 70 K. We used the correction for the sample displacement $\Delta 2\theta \sim \cos(\theta)$, which is valid strictly speaking only in symmetrical diffractions. Therefore there is a shift between fit and measured data for (0 0 L) diffraction for lower L.

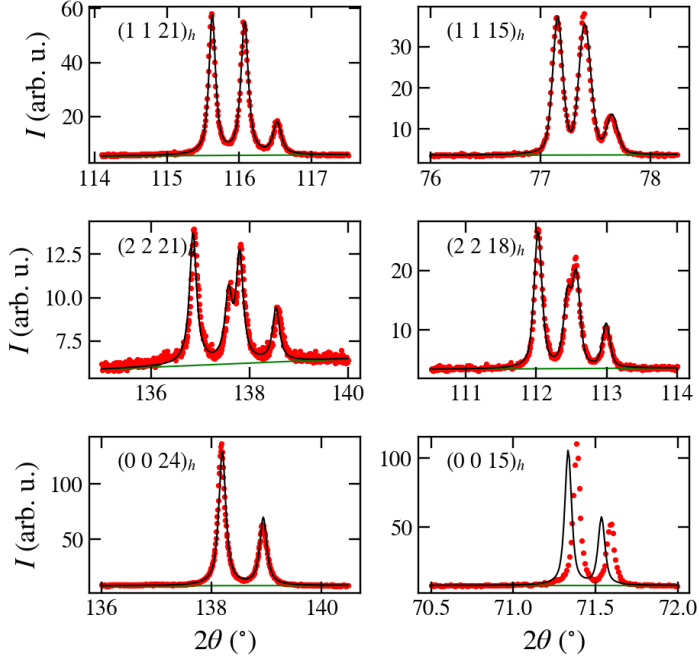

FIG. S12. The comparison of measured data and fits for (2 -1 L) and (4 -2 L) (L = 21, 18, 15) diffractions at 70 K. We used the correction for the sample displacement $\Delta 2\theta \sim \cos(\theta)$, which is valid strictly speaking only in symmetrical diffractions. Therefore there is a shift between fit and measured data for (0 0 L) diffraction for lower L.

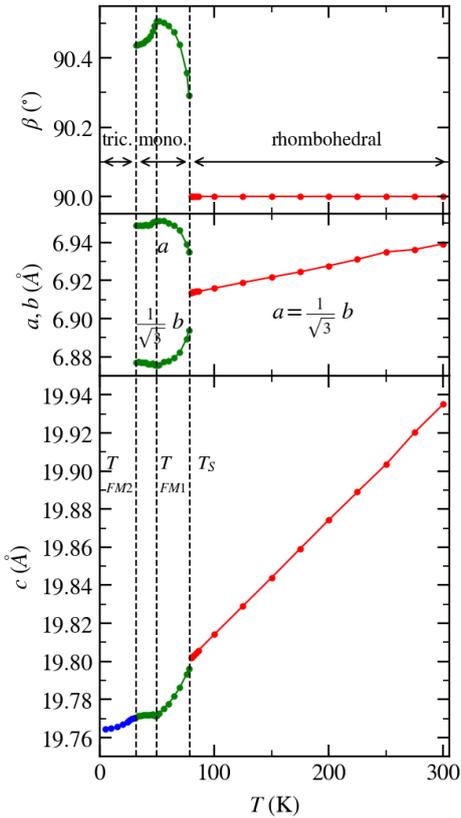

FIG. S13. Temperature dependence of the lattice parameters of VI$_3$.

The possible relation between the lattice distortions due to the magnetic ordering below magnetic structure and the symmetry of magnetic ordering in VI$_3$ can be within the present state of knowledge commented as follows. The magnetization the magnetization curves measured at ~ 2 K in fields parallel and perpendicular to the c-axis, respectively, and presented in papers:

Suhan Son et al., *Phys. Rev.* **B 99**, *041402(R) (2019)*.

Tai Kong et al., *Adv. Mater. 1808074 (2019)*.

J. Yan et al., *Phys. Rev. B 100, 094402 (2019)*.

may be understood when the V magnetic moments are canted or have a small antiferromagnetic component. The *c-a* polar diagram measured at 2 K presented in the latter paper can be conceived with the easy-magnetization direction deflected by ≈ 40° from the *c*-axis. We have measured the *c-a* polar diagrams for temperatures up to higher temperatures and found the deflection is somewhat changing with temperature but it remains in the interval between 40 and 50° from up to $T_{FM1}$. This result might be understood in terms of various magnetic structures. A canted collinear ferromagnetic structure is only one of many possibilities. Without some knowledge of the magnetic structure any specific conclusion about relation of the macroscopic magnetization and the distortion of the monoclinic structure as a consequence of the magnetoelastic interaction due to magnetic ordering can be hardly made. If the magnetic structure is collinear the moments are always far away from the c-axis whereas the monoclinic distortion below $T_S$ leads to only 0.5° deflection of the c-axis just above $T_{FM1}$. The estimated changes of the temperature dependences of lattice parameters due to magnetic ordering below $T_{FM1}$ are very subtle:

$\Delta\beta \approx$ - 0.06 °   ….   0.07% of $\beta$

$\Delta a \approx$ - 0.002 Å   ….   0.03% of $a$

$\Delta b \approx$ + 0.001 Å   ….   0.01% of $b$

$\Delta c \approx$ + 0.006 Å   ….   0.03% of $c$.

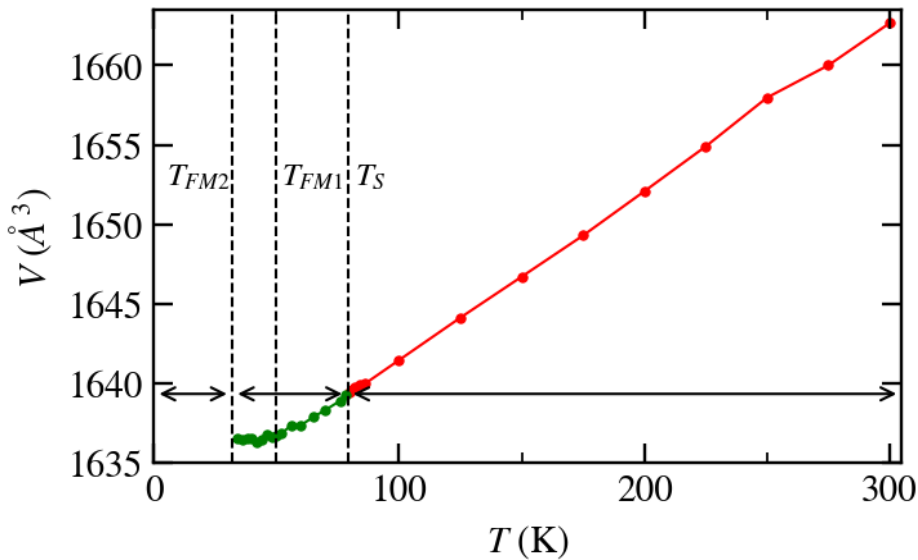

FIG. S14. Temperature dependence of the lattice volume of VI$_3$.

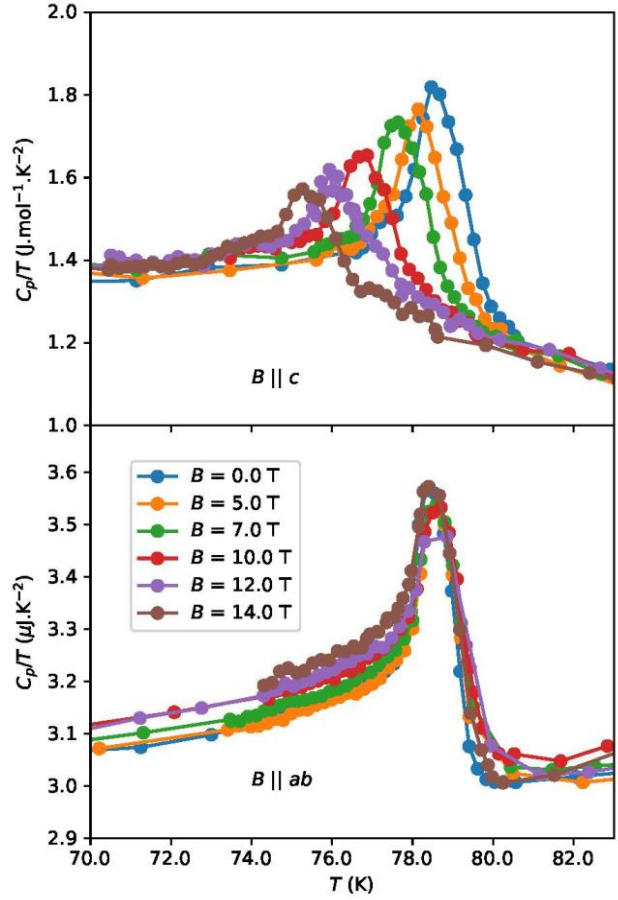

FIG. S15. Temperature dependence of the specific heat ($Cp/T$ vs. $T$ plot) of VI$_3$ in the vicinity of $T_s$, measured in various magnetic fields applied parallel to the $c$-axis (upper panel) and to the $ab$ plane. (lower panel). The two measurements have been performed on different VI$_3$ single-crystal samples in two different experimental setups. Data for $B//ab$ could not be corrected for technical reasons for Apiezon grease on and normalized to sample mass.

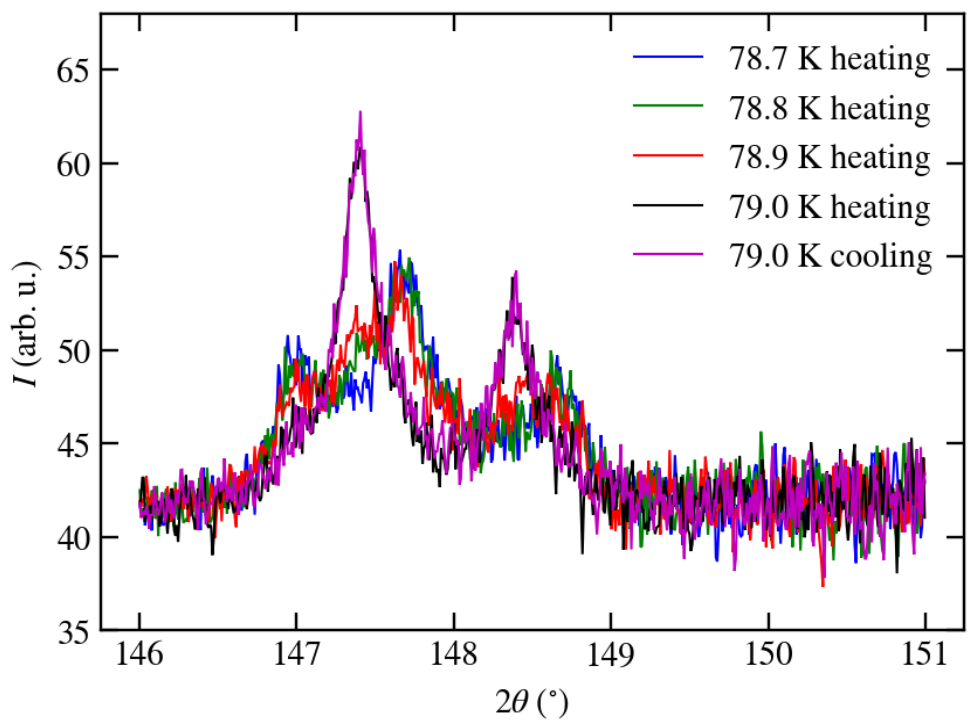

FIG. S16. Thermal cycle around *Ts* of (1 1 24) diffraction peak which shows no hysteresis behavior.